\documentclass{aa}
\usepackage{epsfig,times}
\def\mincir{\raise -2.truept\hbox{\rlap{\hbox{$\sim$}}\raise5.truept \hbox{$<$}\ }}
\def\mincireq{\hbox{\raise0.5ex\hbox{$<\lower1.06ex\hbox{$\kern-1.07em{\sim}$}$}}}
\def\magcir{\raise-2.truept\hbox{\rlap{\hbox{$\sim$}}\raise5.truept \hbox{$>$}\ }}
\def\gr{\kern 2pt\hbox{}^\circ{\kern -2pt K}} 

\def\_{\thinspace}

\begin{document}

\title{
Estimates of relativistic electron and proton energy densities \\
in starburst galactic nuclei from radio measurements
}

\author{Massimo\,Persic\inst{1,2}
	and 
	Yoel\,Rephaeli\inst{3,4}
}

\offprints{M.P.; e-mail: {\tt persic@oats.inaf.it}}

\institute{
INAF-Trieste, via G.B.\,Tiepolo 11, I-34143 Trieste, Italy
         \and
INFN-Trieste, via A.\,Valerio 2, I-34127 Trieste, Italy
        \and
School of Physics \& Astronomy, Tel-Aviv University, Tel Aviv 69978, Israel
	\and
Center for Astrophysics \& Space Sciences,
University of California at San Diego, La Jolla, CA 92093, USA
           }
\date{Received ..................; accepted ...................}

\abstract{The energy density of energetic protons, $U_{\rm p}$, in several nearby 
starburst nuclei (SBNs) has been directly deduced from $\gamma$-ray measurements 
of the radiative decay of $\pi^0$ produced in interactions with ambient protons. 
Lack of sufficient sensitivity and spatial resolution makes this direct deduction 
unrealistic in the foreseeable future for even a moderately distant SBN. A more 
viable indirect method for determining $U_{\rm p}$ in star-forming galaxies is to 
use its theoretically based scaling to the energy density of energetic electrons, 
$U_{\rm e}$, which can be directly deduced from radio synchrotron and possibly also 
nonthermal hard X-ray emission. In order to improve the quantitative basis and 
diagnostic power of this leptonic method we reformulate and clarify its main aspects. 
Doing so we obtain a basic expression for the ratio $U_{\rm p}/U_{\rm e}$ in terms 
of the proton and electron masses and the power-law indices that characterize the 
particle spectral distributions in regions where the total particle energy density 
is at equipartition with that of the mean magnetic field. We also express the field 
strength and the particle energy density in the equipartition region in terms of the 
region's size, mean gas density, IR and radio fluxes, and distance from the observer, 
and determine values of $U_{\rm p}$ in a sample of nine nearby and local SBNs.

}

\maketitle
\markboth{Persic \& Rephaeli: CRs in galaxies}{}

\section{Introduction}

Supernovae (SNe) are thought to be the main drivers of particle acceleration via 
the Fermi-I diffusive shock mechanism (e.g., Gaisser 1990). The radiative yields 
of relativistic electrons and protons have been measured in the radio to TeV regions.

Proton interactions with ambient gas protons produce neutral pions ($\pi^0$), whose 
decay into $\gamma$\,rays is the most significant signature of the main component of 
cosmic rays (CRs). Measurement of this emission yields the essential ingredient in 
the energetics of galactic nonthermal particles, and together with synchrotron and 
Compton emissions by relativistic electrons, allows us to relate diverse phenomena 
such as star formation (SF) and SN rates, efficiency of particle acceleration, and 
magnetic field strength. The high SF rates and dense gas in starburst nuclei (SBNs) 
make these regions prime targets for exploring this relation between stellar and 
nonthermal quantities to explore these environments. 

In a SBN with H-nuclei number density $n$ and volume $V$, the integrated hadronic 
$\gamma$-ray emission from $\pi^0$ decay is 
\begin{eqnarray}
L_{\geq \epsilon}^{[q_{\rm p}]} ~\sim~ \int_V g_{\geq \epsilon}^{[q_{\rm p}]} \, 
	n \, U_{\rm p} \, {\rm d}V ~~~~ {\rm s}^{-1} 
\label{eq:hadr_L}
\end{eqnarray}
with the integral emissivity $g_{\geq \epsilon}^{[q_{\rm p}]}$ measured in units of 
photon\,s$^{-1}$(H-atom)$^{-1}$(eV/cm$^3$)$^{-1}$, and $q_{\rm p}$ the spectral 
index of the proton power-law distribution (Drury et al. 1994). The value of $U_{\rm 
p}$ can be determined from the measured value of $L_{\geq \epsilon}^{[q_{\rm p}]}$ 
if $n(r)$ is known. In steady state the spectro-spatial particle distributions can 
be calculated by numerically solving a convection-diffusion equation which includes 
all the relevant energy losses (e.g., Paglione et al. 1996; Torres 2004). Normalizing 
the particle distributions based on their measured radiative yields estimates of their 
entergy densities. In particular, measurement of the $\pi^0$-decay $\gamma$-ray 
emission is the most direct way to determine $U_{\rm p}$. 

Improving the physical basis for a reliable extraction of the proton energy density, 
$U_{\rm p}$, from $\gamma$-ray measurements is well motivated and timely, in light of 
recent satellite and ground-based observations. These include the detection of three 
nearby starburst galaxies, NGC\,253, NGC\,3034 (M\,82), and NGC\,5945, in the GeV 
(Ackermann et al. 2012) and TeV (Acero et al. 2009; Acciari et al. 2009; Lenain et 
al. 2010) regions. Measured fluxes from these galaxies agree with earlier theoretical 
predictions (NGC\,253: Domingo-Santamar\'\i a \& Torres 2005, Rephaeli et al. 2010; 
NGC\,3034: Persic et al. 2008, de Cea et al. 2009; NGC\,4945: Lenain et al. 2010) based 
on convection-diffusion models for proton and electron propagation and energy losses. 
While there were appreciable differences in the models treated in those works, predicted 
values for $U_{\rm p}$ in the SBNs of the three galaxies were around 250\,eV\,cm$^{-3}$.

Even though most of the energetic particle energy is in protons, the level of the unbeamed 
hadronic emission in most SBNs is not high enough to obtain a reliable estimation of $U_{\rm 
p}$ from the measurement of hadronic $\gamma$-ray emission. This is the main reason why, 
with current detector sensitivities, $\gamma$-ray emission was detected (as we 
just noted) in only three nearby starburst galaxies. It is therefore important to reformulate 
the leptonic route for estimating particle energy densities, testing its viability in these 
three starburst galaxies, and applying the insight gained from applications of both methods 
to these nearby galaxies to improve the precision with which $U_{\rm p}$ and $U_{\rm e}$ can 
be determined in SBNs.

Starting with estimates of the duration of a starburst phase, and of the relevant timescales 
for particle acceleration, energy loss, and advection, we show that particle distributions 
can be in steady state in a SBN. We continue with the usual assumptions in order to relate 
the proton and electron densities by charge neutrality, and energy densities through 
equipartition with the mean 
magnetic field. Doing so we reformulate the hadronic and leptonic methods for determining 
particle energy densities from measurements of radio and $\gamma$-ray emission, and refine 
the expressions for the proton-to-electron density and energy density ratios. We then obtain 
an expression for the equipartition magnetic field and particle energy densities in a SBN in 
terms of an assumed (theoretically-based) value of the spectral index of the proton density, 
and basic measured parameters characterizing the region, i.e., size, gas density, IR luminosity, 
radio flux, radio spectral index, and distance from the observer.

In Section 2 we assess the viability of the standard assumption that particle distributions 
attain a steady state, and the likelihood of reaching energy equipartition with the mean 
magnetic field in the SBN. The basic expressions for proton-to-electron (p/e) ratios and 
for $U_{\rm p}$ in terms of the electron synchrotron flux are written in Sect.\,3. In 
Section 4 these expressions are applied to a sample of nearby and local SBNs. We conclude 
with a discussion and a summary in Sect.\,5.

\section{Energy loss and propagation timescales}

The benchmark timescale for enhanced stellar activity in a SBN, i.e., the duration of the starburst 
phase, is typically estimated to be $t_{\rm SB} \sim 10^8$\,yr. This characteristic time sets the 
scale for the assessment of temporal non-variability of particle distributions. Specifically, 
particle densities may attain a steady state if their characteristic acceleration and the weighted 
energy loss (by collisions and through propagation out of the SBN) timescales are considerably 
shorter than $t_{\rm SB}$. If so, and if the acceleration and energy loss timescales are comparable, 
the particle (spectral) densities are in a steady state.

Acceleration in a SN remnant (SNR) by the Fermi-I process occurs on a timescale $t_{\rm acc} 
\equiv E/ \dot{E} = (\Delta E / E)^{-1} \Delta t = \beta^{-1} \Delta t = (30/\beta_{0.033})\, 
\Delta t \sim 3 \times 10^5 $\,yr, where $\Delta t \sim 10^4$\,yr is a typical SNR lifetime, 
and $\beta_{0.033}$ is the speed of the SN shock in units of $0.033\,c = 10^4$\,km\,s$^{-1}$. 
Clearly, particle acceleration to all relevant energies occurs on a relatively short timescale.

Protons lose energy mainly by proton-proton ({\it pp}) interactions and escape out of the SBN, 
on a characteristic timescale $t_{\rm p}^{-1} = t_{\rm pp}^{-1} + t_{\rm out}^{-1}$. The two 
terms on the r.h.s. are the energy-loss timescales for, respectively, {\it pp} interactions, 
$t_{\rm pp} = (\sigma_{\rm pp} c n)^{-1}$ with $\sigma_{\rm pp}$ the corresponding total cross 
section, and particle removal, $t_{\rm out}^{-1} = t_{\rm adv}^{-1} + t_{\rm diff}^{-1}$, with 
$t_{\rm adv}$ the timescale for advection in a large-scale outflow (i.e., a galactic wind) and 
$t_{\rm diff}$ the diffusion timescale. In the energy range $10-10^5$ GeV, it is $\sigma_{\rm pp} 
\sim 50$ mb, so that $t_{\rm pp} \sim 2 \times 10^5 \, \bigl({ n \over 100\,{\rm cm}^{-3} }\bigr)^
{-1}$\, yr. For a homogeneous distribution of SNe in a SBN of radius $r_{\rm s}$, the advection 
timescale for transfer of particles out of the disk mid-plane region in a fast ($v_{\rm adv} \sim 
1000$\,km\,s$^{-1}$) SB-driven wind
\footnote{
	This velocity seems appropriate for NGC\,3034, given the terminal outflow velocity of $1600 - 
	2200$\,km\,s$^{-1}$ deduced by Strickland \& Heckman (2009; see also Chevalier \& Clegg 1985). }
is $t_{\rm adv} = 5 \times 10^4 (r_{\rm s} / 0.2\,{\rm kpc}) (v_{\rm out} / 1000\,{\rm km\,s}^{-1})^{-1}$\,yr. 
Particle diffusion is likely to be dominated by random-walk through the tangled magnetic fields in 
the SBN. In this dense region the field coherence scale is expected to be much smaller than in the 
disk (where the small-scale cellular structure has a coherence scale $\sim 100$ pc), so scaling to a 
value of 3 pc, the characteristic diffusion time of protons in the SBN is $t_{\rm diff} = 1.3 \times 
10^5 (r_{\rm s} / 0.2\,{\rm kpc})^2 (\lambda /3\,{\rm pc})^{-1}$\,yr, a value comparable to $t_{\rm 
adv}$. From $t_{\rm p}^{-1} = t_{\rm pp}^{-1} + t_{\rm out}^{-1}$, we get $t_{\rm p} \sim 3 \times 
10^4$\,yr for fiducial parameter values typical for SBNs.

The electron energy loss timescale is determined by Coulomb, bremsstrahlung, synchrotron and Compton 
processes, and escape out of the SBN, 
$t_{\rm e}^{-1} =  t_{\rm C}^{-1} + t_{\rm br}^{-1} + t_{\rm SC}^{-1}+ t_{\rm out}^{-1}$, 
respectively. Limiting the discussion here to electron energies higher than a threshold value $\gamma_1$ 
below which Coulomb losses become relevant (see Sect.\,4), we can ignore this non-radiative process in 
the estimation of the weighted mean energy loss time. The bremsstrahlung cooling time for electrons 
traveling through ionized gas with number density $n$ is $t_{\rm br} = 4.4 \times 10^{7} n^{-1}$\,yr. 
The synchrotron-Compton time for electrons traversing a region with disordered magnetic field $B$ with 
energy density  $U_{\rm B}=B^{2}/(8 \pi )$, and IR energy density $U_{\rm IR} = L_{\rm IR}/(4 \pi r^2 c)$, 
is $t_{\rm SC} = {3 m_e c \over 4 \sigma_{\rm T}} \gamma^{-1} (U_{\rm B} + U_{\rm ph})^{-1} \simeq 
\gamma^{-1} (U_{\rm B} + U_{\rm ph})^{-1}$\,yr. Under typical SBN conditions, which are fully 
specified in Sect.\,4 (see also Fig.\,1), the weighted energy loss time for high energy electrons 
is $t_{\rm e} \sim 4.3 \times 10^4$\,yr for $\gamma \sim 10^3$. 

Since typical acceleration and energy loss times are much shorter than the SB duration, particle 
spectral densities can attain steady state at levels that are proportional to the respective ratio 
of the energy loss time to the acceleration time. More generally, in a state of  hydrostatic and 
virial equilibrium, it is likely that in the minimum energy configuration energy densities of 
particles and magnetic fields, which are tightly coupled dynamically and energetically in the SN 
environment, are in energy equipartition (e.g., Longair 1981). Generally, 
the denser and more radiatively intense the environment, the tighter is the coupling between all 
degrees of freedom, including nonthermal particles and magnetic fields. Physical processes that 
couple nonthermal particles and thermal gas are Coulomb interactions (i.e., ionization, 
electronic excitations, and bremsstrahlung), Compton scattering, and excitation of magnetic 
turbulence (e.g., Alfv\' en waves) by non-isotropic particle distributions. Particle coupling to 
magnetic fields is particularly strong because of the high field strength in a SBN and 
its disordered morphology, affecting both particle energy distribution and transport 
properties. Under such conditions, particle-field energy equipartition would be expected.

\section{Particle and field energy densities}

In repeated crossings of the shock region, fast electrons and protons in the ambient SNR gas 
gain energy from their initial fiducial kinetic energy $T_0 \simeq 10\,$keV (in the Maxwellian 
tail) to a very high value, O($10^5$) GeV. In the immediate vicinity of the acceleration sites, 
and before energy losses substantially modify their initial distributions, the particles' spectral 
densities are usually assumed to have a power-law (in momentum $p$) form, $N_{\rm j}(p) = N_{0, 
{\rm j}} p^{-q_{\rm j}}$ with $j=e,p$ for, respectively, electrons and protons; in general, 
$q_{\rm e} \neq q_{\rm p}$. If the gas of nonthermal electrons and protons is approximated as 
an electrically neutral plasma, then 
\begin{eqnarray}
n_o ~=~ \int_{T_0}^{\infty} N_{\rm e}(T)\, {\rm d}T ~=~ \int_{T_0}^{\infty} N_{\rm p}(T)\, {\rm d}T \,.
\label{eq:el_neutr}
\end{eqnarray}
The basic energy-momentum relation yields ${\rm d}p/{\rm d}T$$=$$(T/c^2+m)$$(T^2/c^2+2Tm)^{-1/2}$, 
and using $N_{\rm j}(T)$$=$$N_{\rm j}[p(T)] {\rm d}p/{\rm d}T$, an explicit expression for $N_{\rm 
j}(T)$ is obtained (e.g., Schlickeiser 2002): 
\begin{eqnarray}
N_{\rm j}(T) ~=~ {N_{0, {\rm j}} \over c^2} ~(T+m_{\rm j}c^2)~ \bigl({T^2 \over c^2}+2Tm_{\rm j} 
\bigr)^{-(q_{\rm j}+1)/2}\,.
\label{eq:CR_spectr}
\end{eqnarray}
The respective normalization can now be obtained by performing the integration in 
Eq.\,(\ref{eq:el_neutr}):
\begin{eqnarray}
N_{0, {\rm j}} ~=~ n_0 ~(q_{\rm j}-1)~ \bigl[ {T_0^2 \over c^2} +2T_0 m_{\rm j} 
\bigr]^{(q_{\rm j}-1)/ 2}\,.
\label{eq:CRspectr_norm}
\end{eqnarray}

The p/e {\it number} density ratio, $\zeta = N_{\rm p}(T)/N_{\rm e}(T)$, can now be explicitly written 
\begin{eqnarray}
\lefteqn{
\zeta(T; q_{\rm p}, q_{\rm e}) ~=~ 
{ (q_{\rm p}-1) \over (q_{\rm e}-1) } ~ 
   { [T_0^2+2T_0m_{\rm p}c^2]^{q_{\rm p}-1 \over 2} \over [T_0^2+2T_0m_{\rm e}c^2]^
{q_{\rm e}-1 \over 2} } ~ \times 
}
                \nonumber\\
& & {} \times~  
{ T^{-{(q_{\rm p}+1) \over 2}} (T+m_{\rm p}c^2) (T+2m_{\rm p}c^2)^{-{q_{\rm p}+1 \over 2}}
       \over 
     T^{-{(q_{\rm e}+1) \over 2}} (T+m_{\rm e}c^2) (T+2m_{\rm e}c^2)^{-{q_{\rm e}+1 
\over 2}} } \,.
\label{eq:zeta2}
\end{eqnarray}
When it is assumed that $q_{\rm p}=q_{\rm e}=q$ (e.g., at injection), simpler limiting expressions 
for this ratio are obtained (e.g., Schlickeiser 2002): 
\begin{eqnarray}
\zeta(T; q)   ~=~
	\left\{ \begin{array}{ll} 
        1 ~~~  & \mbox{ ~ ... ~  $T \ll m_{\rm e}c^2$}     \\ 
	\propto {\bigl({T \over m_{\rm p}c^2}}\bigr)^{q-1 \over 2} & \mbox{ ~ ... ~  $m_{\rm e}c^2  
\ll T\ll m_{\rm p}c^2 $}  \\ 
        \bigl({m_{\rm p} \over m_{\rm e}}\bigr)^{q-1 \over 2} & \mbox{ ~ ...~  $T \gg m_{\rm p}c^2$ } \,.
\end{array} \right.
\label{eq:zeta3}
\end{eqnarray}

The general expression for the p/e {\it energy} density ratio, 
\begin{eqnarray}
\kappa(T_0; q_{\rm p}, q_{\rm e}) ~=~ 
{\int_{T_0}^{\infty} N_{\rm p}(T)\,T\,{\rm d}T  \over \int_{T_0}^{\infty} N_{\rm e}(T)\,T\,{\rm d}T} \, ,
\label{eq:kappa1}
\end{eqnarray}
can also be written as: 
\begin{eqnarray}
\lefteqn{
\kappa(T_0; q_{\rm p}, q_{\rm e}) ~=~ 
{ (q_{\rm p}-1) \over (q_{\rm e}-1) } ~ 
   { (T_0^2 + 2T_0 m_{\rm p}c^2)^{q_{\rm p}-1 \over 2} \over (T_0^2 + 2T_0 m_{\rm e}c^2)^
{q_{\rm e}-1 \over 2} } 
~ \times 
} 
                \nonumber\\
& & {} \times ~ 
{ \int_{T_0}^{\infty} T^{-{q_{\rm p}-1 \over 2}} 
                        (T+2m_{\rm p}c^2)^{-{q_{\rm p}+1 \over 2}}
			(T+m_{\rm p}c^2) 
			{\rm d}T 
       \over 
    \int_{T_0}^{\infty} T^{-{q_{\rm e}-1 \over 2}} 
                        (T+2m_{\rm e}c^2)^{-{q_{\rm e}+1 \over 2}}
			(T+m_{\rm e}c^2) 
			{\rm d}T } \,. 
\label{eq:kappa2}
\end{eqnarray}
In order to explore the relevant range of values of this ratio, we computed $\kappa$ 
for several representative values of $q_{\rm p}$ and $q_{\rm e}$ (see Table\,1). An 
approximate expression for $\kappa$ can be obtained by considering only proton and 
electron energies higher than the respective particle mass (where the particle spectra,
$N_{\rm j}(T)$, are single-power-law in energy):
\begin{eqnarray}
\lefteqn{
\kappa(q_{\rm p}, q_{\rm e}) ~\simeq ~
{ (q_{\rm p}-1) \over (q_{\rm e}-1) } ~
{ (q_{\rm e}-2) \over (q_{\rm p}-2) } ~ 
{ (2T_0m_{\rm p}c^2)^{q_{\rm p}-1 \over 2} \over (2T_0m_{\rm e}c^2)^{q_{\rm e}-1 \over 2} } ~
\times} 
                \nonumber\\
& & {}  \times~ 
{ (m_{\rm p}c^2)^{2-q_{\rm p}} \over (m_{\rm e}c^2)^{2-q_{\rm e}} }\,,
\label{eq:kappa3}
\end{eqnarray}
which for $q_{\rm p}=q_{\rm e}=q$ reduces to 
\begin{eqnarray}
\kappa(q) ~ \simeq ~ 
\biggl( {m_{\rm p}  \over m_{\rm e}} \biggr)^{(3-q)/2} 
\,.
\label{eq:kappa4}
\end{eqnarray}
This expression for the energy density ratio (Eq. \ref{eq:kappa4}) is analogous to the high-energy 
limit of the number density ratio in Eq.\,(\ref{eq:zeta3}) (see also Persic \& Rephaeli 2014).


\begin{table*}
\caption[] {Proton-to-electron energy density ratios, $\kappa$$^{[a]}$.}
\begin{flushleft}
\begin{tabular}{ l  l  l  l  l}
\hline
\hline
\noalign{\smallskip}

$\overline{~q_{\rm p} ~~~~q_{\rm e} ~~~~~~~~~~~~~~\kappa~~~~~~~}$ &
~~~~~$\overline{~q_{\rm p} ~~~~q_{\rm e} ~~~~~~~~~~~~~~\kappa~~~~~~~}$ &
~~~~~$\overline{~q_{\rm p} ~~~~q_{\rm e} ~~~~~~~~~~~~~~\kappa~~~~~~~}$ &
~~~~~$\overline{~q_{\rm p} ~~~~q_{\rm e} ~~~~~~~~~~~~~~\kappa~~~~~~~}$ &
~~~~~$\overline{~q_{\rm p} ~~~~q_{\rm e} ~~~~~~~~~~~~~~\kappa~~~~~~~}$ \\

\noalign{\smallskip}
\hline
\noalign{\smallskip}

 2.0 ~~2.0 ~~~~0.258E+02 & ~~~~~2.1 ~~2.0 ~~~~0.984E+01 & ~~~~~2.2 ~~2.0 ~~~~0.418E+01 & ~~~~~2.3 ~~2.0 ~~~~0.197E+01 & ~~~~~2.4 ~~2.0 ~~~~0.101E+01 \\
 2.0 ~~2.1 ~~~~0.628E+02 & ~~~~~2.1 ~~2.1 ~~~~0.239E+02 & ~~~~~2.2 ~~2.1 ~~~~0.102E+02 & ~~~~~2.3 ~~2.1 ~~~~0.479E+01 & ~~~~~2.4 ~~2.1 ~~~~0.246E+01 \\
 2.0 ~~2.2 ~~~~0.119E+03 & ~~~~~2.1 ~~2.2 ~~~~0.453E+02 & ~~~~~2.2 ~~2.2 ~~~~0.193E+02 & ~~~~~2.3 ~~2.2 ~~~~0.906E+01 & ~~~~~2.4 ~~2.2 ~~~~0.466E+01 \\
 2.0 ~~2.3 ~~~~0.189E+03 & ~~~~~2.1 ~~2.3 ~~~~0.720E+02 & ~~~~~2.2 ~~2.3 ~~~~0.306E+02 & ~~~~~2.3 ~~2.3 ~~~~0.144E+02 & ~~~~~2.4 ~~2.3 ~~~~0.740E+01 \\
 2.0 ~~2.4 ~~~~0.269E+03 & ~~~~~2.1 ~~2.4 ~~~~0.102E+03 & ~~~~~2.2 ~~2.4 ~~~~0.436E+02 & ~~~~~2.3 ~~2.4 ~~~~0.205E+02 & ~~~~~2.4 ~~2.4 ~~~~0.105E+02 \\
 2.0 ~~2.5 ~~~~0.357E+03 & ~~~~~2.1 ~~2.5 ~~~~0.136E+03 & ~~~~~2.2 ~~2.5 ~~~~0.578E+02 & ~~~~~2.3 ~~2.5 ~~~~0.272E+02 & ~~~~~2.4 ~~2.5 ~~~~0.140E+02 \\
 2.0 ~~2.6 ~~~~0.451E+03 & ~~~~~2.1 ~~2.6 ~~~~0.172E+03 & ~~~~~2.2 ~~2.6 ~~~~0.731E+02 & ~~~~~2.3 ~~2.6 ~~~~0.344E+02 & ~~~~~2.4 ~~2.6 ~~~~0.177E+02 \\
 2.0 ~~2.7 ~~~~0.551E+03 & ~~~~~2.1 ~~2.7 ~~~~0.210E+03 & ~~~~~2.2 ~~2.7 ~~~~0.892E+02 & ~~~~~2.3 ~~2.7 ~~~~0.420E+02 & ~~~~~2.4 ~~2.7 ~~~~0.216E+02 \\
 2.0 ~~2.8 ~~~~0.654E+03 & ~~~~~2.1 ~~2.8 ~~~~0.249E+03 & ~~~~~2.2 ~~2.8 ~~~~0.106E+03 & ~~~~~2.3 ~~2.8 ~~~~0.499E+02 & ~~~~~2.4 ~~2.8 ~~~~0.256E+02 \\
 2.0 ~~2.9 ~~~~0.760E+03 & ~~~~~2.1 ~~2.9 ~~~~0.289E+03 & ~~~~~2.2 ~~2.9 ~~~~0.123E+03 & ~~~~~2.3 ~~2.9 ~~~~0.579E+02 & ~~~~~2.4 ~~2.9 ~~~~0.298E+02 \\
 2.0 ~~3.0 ~~~~0.867E+03 & ~~~~~2.1 ~~3.0 ~~~~0.330E+03 & ~~~~~2.2 ~~3.0 ~~~~0.140E+03 & ~~~~~2.3 ~~3.0 ~~~~0.661E+02 & ~~~~~2.4 ~~3.0 ~~~~0.340E+02 \\

\noalign{\smallskip}
\hline
\end{tabular}
\end{flushleft}
\smallskip

\noindent
$^{[a]}$ An energy range of $10^{-5}\,{\rm GeV} - 10^{5}\,{\rm GeV}$ is assumed.
\smallskip
\end{table*}

Determining $U_{\rm p}$ from the theoretically predicted value of $\kappa$ and $U_{\rm e}$, which is 
deduced from the measured radio flux, obviously requires knowledge of the mean magnetic field in the 
emitting region, $B$. To overcome this (implied) indeterminacy, the assumption of field and particle 
energy equipartition is commonly made. In addition, the contribution of secondary electrons (from $\pi^-$
decay)
\footnote{
	Secondary positrons (from $\pi^+$decay) almost immediately annihilate 
	with thermal electrons.
	}
to the (steady state) electron density has to be included. 

While the exact form of the particle steady-state spectral density does not generally have a single 
power-law form (e.g., Rephaeli 1979), the radiative yields are largely by protons and electrons with 
energies higher than a few Gev, for which Coulomb losses (which flatten the spectral density) are 
subdominant. In this limit, the total electron spectral density can be approximated by
\begin{eqnarray}
N_{\rm e}(\gamma) ~=~ N_{e,{\rm 0}}\, (1+\chi) ~ \gamma^{-q_{\rm e}} , & & 
\label{eq:el_spectrum}
\end{eqnarray}
where the electron Lorentz factor $\gamma$ is in the range $\gamma_1 \leq \gamma \leq \gamma_2$, $N_
{e,{\rm 0}}$ is a normalization factor of the primary electrons, and $\chi$ is the secondary-to-primary 
electron ratio. The electron spectral index is $q_{\rm e} \geq 2$, with the minimal value of 2 
corresponding to the strong-shock limit of the Fermi-I acceleration mechanism.

Ignoring the contribution of low-energy electrons with $\gamma < \gamma_1$, the electron energy density 
is $U_{\rm e} \simeq N_{\rm e,0}\,(1+\chi)\,m_{\rm e} c^2\, \int_{\gamma_1}^{\gamma_2} \gamma^{1-q_{\rm e}} 
{\rm d}\gamma$, where $\gamma_2$ is the upper end of the electron energy spectrum. For $q_{\rm e}>2$, 
\begin{eqnarray}
U_{\rm e} = N_{e,{\rm 0}} ~(1+\chi)  ~m_{\rm e}c^2  ~\gamma_1^{2-q_{\rm e}} {[1-
(\gamma_2/\gamma_1)^{2-q_{\rm e}}] \over (q_{\rm e}-2)}\,. 
\label{eq:el_en_dens1}
\end{eqnarray}

For a population of electrons described by Eq.\,(\ref{eq:el_spectrum}), traversing a homogeneous
magnetic field of strength $B$ that permeates a region of (spherically equivalent) radius $r_{\rm 
s}$ located at a distance $d$ from the observer, and emitting a 5\,GHz synchrotron radiation flux 
of $f_5$ Jy, the standard synchrotron formula (e.g., Blumenthal \& Gould 1970) yields 
\begin{eqnarray}
N_{\rm e,0} (1+\chi) ~=~ 1.6 \times 10^{-16} \,a_{q_{\rm e}}^{-1} \, \psi_5 \, 1250^{q_{\rm e} 
\over 2}\,B^{-{q_{\rm e}+1 \over 2}} \,,
\label{eq:el_spect_norm}
\end{eqnarray}
where quantities are expressed in c.g.s. units, the factor $a_{q_{\rm e}}$ is defined and tabulated 
(in, e.g., Blumenthal \& Gould 1970), and $\psi_5 \equiv ({r_{\rm s} \over 0.1\,{\rm kpc}})^{-3} (
{d \over {\rm Mpc}})^2 ({f_5 \over {\rm Jy}})$. From Eqs.\,(\ref{eq:el_en_dens1}) and 
(\ref{eq:el_spect_norm}) we derive
\begin{eqnarray}
\lefteqn{
U_{\rm e} ~=~ 1.3 \times 10^{-22} \, 1250^{q_{\rm e} \over 2} \, \psi_5 ~ 
B^{-{q_{\rm e}+1 \over 2}} ~\times
} 
                \nonumber\\
& & {} ~~~~~~~~~~~~~~~~ \times~
{\gamma_1^{2-q_{\rm e}} [1-(\gamma_2/\gamma_1)^{2-q_{\rm e}}]
\over (q_{\rm e}-2)~ a_{q_{\rm e}} } \, 
 ~~~~ {\rm erg\,cm}^{-3}\,. 
\label{eq:el_en_dens2}
\end{eqnarray}

Since $U_{\rm e}$ includes both primary and secondary electrons, the rough assumption that both 
populations can be characterized with nearly the same power-law index
\footnote{
	Theoretical expectation is that power-law indices of proton and electron source 
	spectra are nearly identical. Under typical interstellar medium (ISM) conditions, 
	electrons 
	lose energy more efficiently than protons, resulting in a relative steepening of 
	the electron spectrum. Secondary electrons (and positrons, produced in $\pi^\pm$ 
	decays following {\it pp} collisions) initially have a slightly flatter spectrum (by 
	$\Delta q \simeq 0.2$) than the parent proton spectrum for energies $\magcir$1\,GeV, 
	but then their spectrum 
	steepens due to severe energy losses. Detailed numerical models of emission from 
	starburst galaxies show that primary and secondary electrons have roughly similar 
	spectral shapes at energies $\sim 10$ MeV -- $1$ TeV (e.g., Domingo-Santamar\'\i a 
	\& Torres 2005; De Cea del Pozo et al. 2009; Rephaeli et al. 2010). Therefore, in 
	the relatively small SBN region, where energetic particles have not yet ventured out 
	too far from their acceleration sites, it is quite reasonable (for our purposes 
	here) to characterize their spectra with the same index. After all, from a practical 
	point of view, both primary and secondary electrons contribute to the main observable 
	upon which we base our 	analysis - radio emission - which is described by a single-index 
	power-law spectrum.}
means that the primary electron energy is $U_{\rm e}/(1+\chi)$. Denoting the primary p/e energy 
density ratio (see Sect.\,3) by $\kappa(q_{\rm p}, q_{\rm e})$, the proton energy density is 
\begin{eqnarray}
U_{\rm p} ~\simeq~ \kappa(q_{\rm p}, q_{\rm e}) ~{U_{\rm e} \over 1+\chi}  \,.
\label{eq:prot_en_dens}
\end{eqnarray}
Since tight coupling is expected in the very dense environment of SBN, particle and magnetic field 
energy densities can be assumed to be close to equipartition (see Eq.\,(\ref{eq:field_part_equil})). 
If so, we can express the field in terms of the total particle energy density; this leads to 
\begin{eqnarray}
\lefteqn{ 
U_{\rm p}  \, = \, {2.5 \times 10^{10} \over 1\,+\, (1+\chi)/ \kappa } ~
\biggl[ 3.3 \times 10^{-21} \, \biggl(1+{\kappa \over 1+\chi}\biggr)\, \gamma_1^{2-q_{\rm e}}\, 
\times 
 }    
	\nonumber\\
 & & {}  \times\,
{[1-(\gamma_2/\gamma_1)^{2-q_{\rm e}}] \, 1250^{q_{\rm e}/ 2} \, 
\psi_5 \over (q_{\rm e}-2)\, a_{q_{\rm e}} } \biggr]^{4/(5+q_{\rm e})} \, 
{\rm eV\,cm}^{-3}\,. 
\label{eq:U_p}
\end{eqnarray}
In general, $q_{\rm e}$, $q_{\rm p}$, $\chi$, $\gamma_1$, $\gamma_2$, and $\kappa$ need to 
be known (or assumed) in order to compute $U_{\rm p}$. The value of $q_{\rm e}$ is readily deduced 
from measurements of the 
(nonthermal) radio spectral index, $\alpha$, through the relation $q_{\rm e}= 2\, \alpha + 1$.  

\begin{figure}
\vspace{8.0cm}
\includegraphics{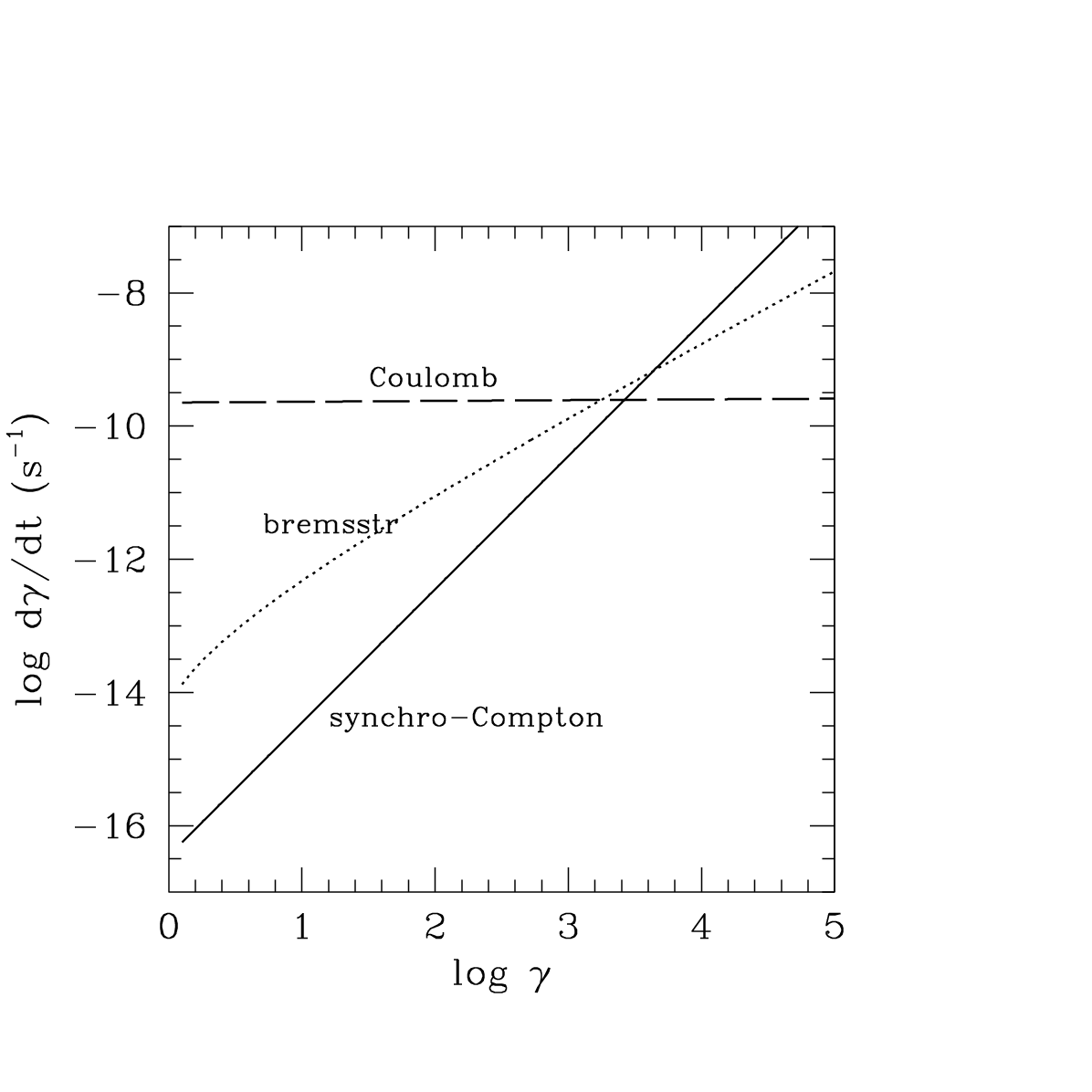}
\caption{Energy loss rates of an electron of energy $\gamma\, m_{\rm e} c^2$ due to Coulomb, 
bremsstrahlung, and synchrotron-Compton processes in a typical (M\,82-like) SBN environment: 
$B=100\,\mu$G, $L_{\rm IR} = 10^{44}$\,erg\,s$ ^{-1}$, $r_{\rm s} = 0.2$\,kpc, $n=100$\,cm$^{-3}$, 
$n_e=200$\,cm$^{-3}$. 
}
\label{fig:en_loss}
\end{figure}


\subsection{Proton spectral index}

The proton spectral index has been measured in the nearby starburst galaxies NGC\,253, NGC\,3034, 
and NGC\,4945, as $q_{\rm p} \simeq 2.2$ (Ackermann et al. 2012). As we discuss below, it can be 
expected that this value also characterizes proton spectra in other starburst galaxies.

Suprathermal particles injected into a supernova shock have a power-law spectrum with index 
$q=(R+2)/(R-1)$, where $R$ is the shock compression ratio, defined as the downstream to upstream 
density, $\rho_d/\rho_u$. In an ideal gas, $R = (\gamma+1)/[(\gamma-1)+2/M_u^2]$, where $\gamma$ 
is the ratio of the gas specific heats, $M_u = v_u/c_{\rm s}$ is the shock Mach number, and $c_{\rm s}$ 
is the sound speed. Since $c_{\rm s} = \sqrt{\gamma p_u/\rho_u}$, clearly $M_u = v_u/\sqrt{\gamma 
k_{\rm B}T_u/(\mu m_{\rm p})}$. The temperature of the upstream medium, $T_u$, is 
nearly two orders of magnitude higher ($T \sim 10^{6-7}\gr$) in SB regions than in more quiescent 
galactic disks ($T \sim 10^{4-5}\gr$) (Heckman \& Lehnert 2000; Fujita et al. 2009). Therefore, for a 
given shock velocity $v_u$, during the Sedov phase of SNR the Mach number is smaller in the SBN than in 
the disk, so that the compression ratio of a strong shock in the SBN is expected to be $R \simeq 3.6$, 
somewhat lower than the more typical (Galactic) value $R \simeq 4$. This lower value implies an injection 
index in the range $2.0 \leq q \leq 2.3$ (Fujita et al. 2009).

\subsection{Electron secondary-to-primary ratio}

The secondary-to-primary electron ratio $\chi$ depends on the injection p/e number ratio, $\zeta$, 
and on the gas density, which determines the effectiveness of {\it pp} interactions that yield 
charged and neutral pions. An electron is produced in the decay $\pi^- \rightarrow \mu^- + 
\bar\nu_\mu$, followed by $\mu^- \rightarrow e^- + \bar\nu_e + \nu_\mu$; a positron is produced 
in the decay $\pi^+ \rightarrow \mu^+ + \nu_\mu$, followed by $\mu^+ \rightarrow e^+ + \nu_e + 
\bar\nu_\mu$. The {\it pp} branching ratios are such that in 2/3 of these interactions $e^\pm$ 
are produced.
 
The mean free path of an energetic proton for {\it pp} interactions in a gas with density $n$ is 
$\lambda_{\rm pp} = (\sigma_{\rm pp} n)^{-1}$. The {\it pp} cross section for protons with kinetic 
energy of a few TeV is $\sigma_{\rm pp} \approx 50 \,{\rm mb} = 5 \times 10^{-26}$\,cm$^2$ 
(Baltrusaitis et al. 1984), so that the probability for a proton to undergo {\it pp} interactions 
during its 3D random walk through a region of radius $r_{\rm s}$ is then $\sqrt{3} \,r_{\rm s} / 
\lambda_{\rm pp}$. Given the injection p/e ratio, $\zeta$, and the above branching ratio, the 
secondary-to-primary electron ratio is 
\begin{eqnarray}
\chi ~=~ {2 \over 3} ~ \zeta\, ~  \sqrt{3} ~ r_{\rm s} ~ n ~ \sigma_{\rm pp} \,.
\label{eq:chi}
\end{eqnarray}
In a typical SBN with $r_{\rm s} = 0.2$\,kpc, $n = 200$\,cm$^{-3}$, and $q = 2.2-2.3$), estimated 
values of $\chi$ are $\sim 0.6-1$, in agreement with results from more detailed numerical models 
(e.g., Paglione et al. 1996; Domingo-Santamar{\'\i}a \& Torres 2005; De Cea et al. 2009; Rephaeli 
et al. 2010).

\subsection{Equipartition magnetic field}

As we have already noted, it is quite likely that in their equilibrium minimum-energy configuration 
particles and magnetic fields, which are tightly coupled dynamically and energetically in the SN 
environment, are in energy equipartition. We assume that equipartition is indeed attained during the 
starburst phase, and use it to determine the mean magnetic field in the SBN region. The starting 
point is an estimate of the electron energy density from the measured radio flux, which we obtain 
by integrating the electron spectral energy density over the interval $[\gamma_1\, , \gamma_2]$.  

For consistency with the assumed power-law form of the electron spectral density, we take the 
low-energy limit $\gamma_1$ to be the value of the Lorentz factor at which the sum of the Coulomb 
(or electronic excitation, in ionized gas) and bremsstrahlung loss rates equals the synchrotron-Compton 
loss rate. This is also based on the fact that even for the relatively high values of the magnetic 
field in SBN, the measured radio emission (upon which our normalization of the electron density is 
based) samples electrons with $\gamma > 10^3$. 

The Coulomb (electronic excitation) loss rate (e.g., Rephaeli 1979) is 
\begin{eqnarray}
b_0 ~=~ - \dot \gamma_{\rm c} ~\simeq~
1.2 \times 10^{-12} \, n_e \, \biggl[ 1.0 +{{\rm ln}\,(\gamma/n_e) \over 84} \biggr] ~
{\rm s}^{-1}\,,
\label{eq:coul_loss}
\end{eqnarray}
where $n_e$ is the (thermal) electron number density. 

The closely related bremsstrahlung loss rate for a H+He plasma with solar abundances, that also includes 
the contribution of e-e scatterings, is (Gould 1975) 
\begin{eqnarray}
b_1 = - \dot \gamma_{\rm b} \simeq 
	\left\{ \begin{array}{ll} 
	1.78 \times 10^{-16} n \gamma \bigl[ {\rm ln}\,(\gamma) + 0.36 \bigr] ~ {\rm s}^{-1}  & \mbox{ ion. }\\
        9.44 \times 10^{-16} n \gamma ~ {\rm s}^{-1}  & \mbox{ neutr.}\,,
\end{array} \right.
\label{eq:bremss_loss}
\end{eqnarray}
where $n$ denotes the number density of hydrogen nuclei. The reported expression for neutral plasma 
holds for $\gamma \geq 100$; at lower $\gamma$ the ionized and neutral cases essentially overlap (Gould 1975).

The higher order (in $\gamma$) synchrotron-Compton loss rate (e.g., Blumenthal \& Gould 1970) is  
\begin{eqnarray}
b_2 ~=~ -\dot \gamma_{\rm SC} ~=~  
1.3\times 10^{-9} \, \gamma^2\, \bigl( B^2 + 8\pi \rho_{\rm IR} \bigr)~
{\rm s}^{-1}\,,
\label{eq:syn_loss}
\end{eqnarray}
where 
$\rho_{\rm IR}$ is the energy density of the (dominant) IR radiation field in the SBN region. 

Equating the 
sum of the first two loss rates with the latter yields an estimate of $\gamma_1$. In Fig.\,\ref{fig:en_loss} 
we display the energy-loss rates, expressed in Eqs.\,(\ref{eq:coul_loss})-(\ref{eq:syn_loss}), for typical SBN 
parameters. In our numerical estimates (see Table\,2) we use the second of Eq.\,(19) for the mostly-neutral 
SBNs in Arp\,220 and Arp\,299-A, and the first for the other, mostly-ionized SBNs. 

Whereas the dependence of  the electron energy density on $\gamma_1$ is appreciable, since 
$\gamma_{2} >10^5$ the exact value of the upper end of the $\gamma$ integral is of little significance 
for the range of values of $q_e$ of interest here; in our calculations we take $\gamma_2 = 10^5$.
  
With both $U_{\rm e}$ and $U_{\rm p}$ determined, the mean field strength is deduced from
\begin{eqnarray}
{B^2 \over 8 \pi} ~=~ \eta ~ (U_{\rm p} + U_{\rm e})\,, 
\label{eq:field_part_equil}
\end{eqnarray}
where $\eta=1$ in equipartition, but somewhat lower, $\eta=3/4$, in strictly minimum-energy configuration 
(Longair 1981)
\footnote{
	Energy equipartition may be attained (over time) under conditions of tight coupling of the 
	main matter and energy components of a physical environment in its minimum energy configuration.
	The electromagnetic emission of a SBN that directly comes from relativistic particles manifests
	itself in the radio and in $\gamma$\,rays. The total energy density in the SBN responsible for 
	such nonthermal emission is $U=U_{\rm CR} +U_{\rm B}$. As noted in Sect.\,2, under physical 
	conditions prevailing in a SBN the particles and the B-field behave like relativistic fluids 
	tightly coupled with each other. Their equilibrium configuration most likely corresponds to a 
	state of minimum energy, which is achieved for $U_{\rm B}=\eta \,U_{\rm CR}$ with $\eta = 3/4$ 
	(Longair 1981). Clearly, this minimum-energy condition is very close to equipartition ($\eta=1$), 
	so at equilibrium the particle energy density nearly equals the field energy density.
}. 
In terms of the particle p/e energy density ratio, $\kappa$, the particle-field coupling 
condition is $B^2/8\pi = \eta \, U_{\rm p} \, [1+ (1+\chi)/\kappa]$, so that 
\begin{eqnarray}
\lefteqn{
B ~=~ \sqrt{\eta} ~ \biggl[ 3.3 \times 10^{-21} \, \biggl(1+{\kappa \over 
1+\chi}\biggr)  ~ 1250^{q_{\rm e}/ 2}~  \psi_5 \times ~  }
        \nonumber\\
 & & {}  ~ ~ ~ ~ ~  \times ~
\gamma_1^{2-q_{\rm e}} ~ {[1-(\gamma_2/\gamma_1)^{2-q_{\rm e}}]  
\over (q_{\rm e}-2) ~ a_{q_{\rm e}}} \biggr]^{2 / (5+q_{\rm e})} ~ ~ {\rm G}\,.
\label{eq:equip_B}
\end{eqnarray}
Given the values of $n$, the size of the SBN region, and the IR energy density (which is computed from 
the $L_{\rm IR}$ and the region size), as listed in Table\,2, the expression for $B$ in the last equation 
(with $\eta=1$) can be inserted in Eq.\,(\ref{eq:syn_loss}), and using the other two loss rate equations, 
the value of $\gamma_1$ can be deduced. This value of $\gamma_1$ is then substituted in the equation for 
$B$ to obtain the equipartition value of the mean field. We note that in most cases the corresponding 
synchrotron frequency, $\nu_1 = 4 B \gamma_1^2$\,MHz, where the radio spectrum is expected to curve to 
a flatter slope for decreasing frequencies, is comparable to our reference frequency of 5\,GHz.

\section{Proton energy densities in SBNs}

In starburst galaxies SF is very intense in a relatively small nuclear region (the SBN) with effective 
radius $r_{\rm s} \approx 0.2$\,kpc -- with $r_{\rm s} \equiv (3/4 R_{\rm sb}^2 h_{\rm sb})^{1/3}$, where 
$R_{\rm sb}$ and $h_{\rm sb}$ are the radius and height of the region. By contrast, low-intensity star 
formation quiescently proceeds throughout the galactic disk (as in normal, i.e., non-SB, spirals, such as 
the Galaxy).


\begin{table*}
\caption[] {Starburst galaxies: data and results for high-energy activity.}
\begin{flushleft}
\begin{tabular}{ l  l  l  l  l  l  l  l  l  l  l  l  l  l  l}
\hline
\hline
\noalign{\smallskip}

Object                      & 
$~D_L^{[1]}$                & 
$~R_{\rm SB}^{[2]}$        & 
$h_{\rm SB}^{[3]}$          & 
$~f_5^{[4]}$                & 
$~\alpha^{[5]}$             & 
$~~n_e^{[6]}$               &
$~L_{\rm IR}^{[7]}$         &
$M_{\rm SB}^{[8]}$          &
$~\chi^{[9]}$             &
$~~~\kappa^{[10]}$          &
$~~\gamma_1^{[11]}$         &
$B^{[12]}$                  &
$U_{\rm p}^{[13]}$        &
~~Notes$^{[a]}$                       \\
\noalign{\smallskip}
\hline
\noalign{\smallskip}
Arp\,220 E & 74.7 &  114$^+$ &~~ --        & 0.08 & 0.70& 3000$^+$& 44.91& 9.3  &  24 &~~48  &  21000  & 155  & 390 & \\
Arp\,220 W & 74.7 & ~~70$^+$ &~~ --        & 0.10 & 0.70& 3000$^+$& 45.08& 9.1  &  40 &~~48  &  15000  & 230  & 730 & \\
Arp\,299-A & 43.0 &  140     & 200$^\star$ & 0.10 & 0.60& ~    250& 44.88& 9.0  & ~~8 &~~20  & ~~8700  & 145  & 365 & = IC \,694\\
NGC\,~~253 &~~2.5 &  180     & 150         & 1.80 & 0.70& ~    100& 43.62& 7.7  & 0.3 &~~48  & ~~8300  & 100  & 235 & \\
NGC\,3034  &~~3.4 &  300     & 200         & 3.70 & 0.71& ~    200& 43.96& 8.1  & 0.3 &~~51  & ~~6600  & 100  & 250 & = M\,82\\
NGC\,3628  &~~7.6 &  135     & 200$^\star$ & .065 & 0.86& ~    100& 43.30& 7.3  & 0.1 & 120  & ~~7600  &~~65  & 100 & \\
NGC\,4945  &~~3.7 &  250     & 200$^\star$ & 2.25 & 0.60& ~    300& 43.72& 7.4  & 0.1 &~~20  & ~~4700  & 110  & 270 & \\
NGC\,5236  &~~3.7 &  180     & 200$^\star$ & 0.75 & 0.80& ~    200& 43.45& 7.3  & 0.1 &~~90  & ~~5000  & 105  & 260 & = M\,83\\
NGC\,6946  &~~5.5 &  150     & 200$^\star$ & .045 & 0.74& ~    100& 43.51& 7.0  & 0.7 &~~60  & ~~4000  &~~65  & 110 & \\

\noalign{\smallskip}
\hline
\end{tabular}
\end{flushleft}
\smallskip

\noindent
{\small
$^{[1]}$Distance, in Mpc (from Ackermann et al. 2012).
$^{[2,3]}$Radius and height of (cylindrical) star-forming region, in pc ($\star$: assumed; $+$: spherical). 
Data are from Torres 2004 and references therein (Arp\,220), Sugai et al. 1999 (Arp\,299-A), 
Carral et al. 1994 (NGC\,253), V\"olk et al. 1996 (NGC\,3034), Israel et al. 1990 (NGC\,3628), 
Brock et al. 1988 (NGC\,4945), Calzetti et al. 1999 (NGC\,5236), Schinnerer et al. 2006 (NGC\,6946).
$^{[4,5]}$Nonthermal 5\,GHz flux density, in Jy, and spectral radio index. Data are from Torres 
2004 (Arp\,220), Bondi et al. 2012 (Arp\,299-A), Rephaeli et al. 2010 and references therein (NGC\,253), 
Klein et al. 1988 and Carlstrom \& Kronberg 1991 (NGC\,3034), Condon et al. 1982 (NGC\,3628), 
Elmouttie et al. 1997 (NGC\,4945), Sukumar et al. 1987 (NGC\,5236), Murphy et al. 2011 (NGC\,6946).
$^{[6]}$Thermal electron density, in cm$^{-3}$ ($\star$: assumed). Data are taken or computed from 
Yun et al. 2004 (Arp\,220), Sugai et al. 1999 and Zhao et al. 1997 (Arp\,299-A), 
Engelbracht et al. 1998 (NGC\,253), Petuchowski et al. 1994 (NGC\,3034), Dahlem et al. 
1996 (NGC\,3628), Lipari et al. 1997 and Spoon et al. 2000 (NGC\,4945), Krabbe et 
al. 2014 (NGC\,5236), Engelbracht et al. 1996 (NGC\,6946). 
$^{[7]}$Log of the total IR (8--1000$\mu$m) luminosity (erg\,s$^{-1}$; Ackermann et al. 2012, and 
Charmandaris et al. 2002 for Arp\,299-A). 
$^{[8]}$Log of ISM mass in the SB (in $M_\odot$). Data are from Torres 2004 (Arp\,220), 
Charmandaris et al. 2002 (Arp\,299-A), 
Carral et al. 1994 and Domingo-Santamar\'\i a \& Torres 2005 (NGC\,253), 
Rickard et al. 1977, Rieke et al. 1980, Weiss et al. 2001 (NGC\,3034), 
Israel et al. 1990 (NGC\,3628), Spoon et al. 2000 (NGC\,4945), 
Israel \& Baas 2001 (NGC\,5236, NGC\,6946). 
$^{[9]}$Secondary-to-primary electron ratio.
$^{[10]}$p/e energy density ratio. 
$^{[11]}$Electron Lorentz factor at which Coulomb and synchrotron energy losses are equal.
$^{[12]}$Equipartition magnetic field, in $\mu$G. 
$^{[13]}$Equipartition CR proton energy density, in eV\,cm$^{-3}$. 
$^{[a]}$ Arp\,220 E and W denote, respectively, the east and west extreme starbursts 
embedded in Arp\,220's molecular disk (see Torres 2004 and references therein). NGC\,4945 
hosts a Seyfert-1 nucleus, but its high-energy $\gamma$\,ray emission is dominated by the 
SBN (Spoon et al. 2000; Lenain et al. 2010).
}
\end{table*}

For a sample of local SBNs, values of  $U_{\rm p}$ and $B$ were calculated using the relevant 
observational quantities in Table\,2, starting with the radio flux and spectral index. Because 
of appreciable observational uncertainties, mostly in the values of the gas density and the 
size of the emitting region, and because of modeling uncertainties, the derived values of $B$ 
and $U_{\rm p}$ are also uncertain, typically by a factor which we estimate to be $\sim$1.4 
and $\sim$2. While our results for these two quantities would not seem to be that precise, it 
should be emphasized that some of the uncertainties are inherent given the basic difficulties 
in determining the size and density of the SBN, the limited spatial resolution, and the 
rudimentary level of the spectral $\gamma$-ray measurements. We note that, in light of these 
substantial uncertainties, the (modeling) uncertainty in $B$, when the latter is calculated 
assuming energy equipartition as compared to minimum energy, is relatively insignificant. 
For example, for the two well-studied nearby starburst galaxies, NGC\,253 and NGC\,3034 (M\,82), 
values of $B$ are only $\sim 15\%$ lower in the minimum energy configuration compared to the 
equipartition values (listed in Table\,2). The CR-derived results are stable over much of the 
ISM parameter space of SBNs (Boettcher et al. 2013).

Our electron-based estimates of the proton energy densities are in agreement with $\gamma$-ray 
measurements of $\pi^0$-decay emission for the three galaxies NGC\,253, NGC\,3034, and 
NGC\,4945 for which such emission was detected. 
The results for the very compact SBNs of Arp\,220 and Arp\,299-A represent extreme cases; small 
source regions and relatively hard electron spectra result in high magnetic fields and high CR 
energy densities, $B \sim 0.2$\,mG and $U_{\rm p} \sim 500$\,eV\,cm$^{-3}$. However, it is 
questionable whether in such extreme environments, with conditions that are more typical of 
SNRs, steady state and equipartition are actually attained (e.g., Torres 2004). The unusually 
high values of $\gamma_1$ derived for these highly compact nuclei may signal, in fact, a 
breakdown in our assumptions, the most critical one being particle-field equipartition.

\section{Discussion}

Active SF leads to particle acceleration by SN shocks. Considerations of the 
acceleration, relevant energy losses, and starburst timescales, indicate that relativistic 
proton and electron distributions reach steady state during most of the starburst phase. 
Given the tight coupling between the particles and magnetic fields in the dense plasma, it 
is quite likely that energy equipartition is achieved in the SBN region. With an assumed 
theoretically motivated p/e ratio, the assumption of equipartition provides the requisite 
relation to determine particle energy densities and the mean field from spectral radio 
measurements. 

Essential to this radio-based method is a reliable estimate of the p/e energy ratio, $\kappa$. 
Adopting the common assumption of an overall electrically neutral nonthermal plasma, we derived 
an approximate expression for this ratio as a function of the electron and proton spectral 
indices, $q_{\rm e}$ and $q_{\rm p}$. We note that, in the limit of $q_{\rm e}= q_{\rm p}=q$ 
(e.g., at injection), $\kappa \simeq (m_{\rm p}/m_{\rm e})^{(3-q)/2}$; this simple relation is 
supplementary to the analogous, well-known p/e number ratio, $N_{\rm p}/N_{\rm e} = (m_{\rm p}/
m_{\rm e})^{(q-1)/2}$ (e.g., Schlickeiser 2002). Even though the assumptions of single-index 
steady-state spectra and particle-field equipartition may be unrealistic, at our present 
knowledge of the SBN environment relaxing any of these simplifications necessarily leads to 
a more parameter-rich formalism that will invariably result in arbitrariness in guessing 
parameter values in what is essentially an underdetermined problem.

For the determination of $q_{\rm p}$, we have assumed that SBN $\gamma$-ray spectra (in the 
observed {\it Fermi}/LAT and IACT energy ranges) are dominated by emission from $\pi$-decay 
owing to the much higher SF rate and mean (target) gas density in the SBN than in the (rest of 
the) disk. The theoretical expectation of proton injection index in the range $2.0 \leq q \leq 2.3$ 
is fully consistent with the measured value $q_{\rm p}\simeq 2.2$. Because of the lack of spatial 
information on the distribution of $\gamma$-ray emission in the nearby starburst galaxies, our 
expectation on the respective contributions of the SBN and disk regions is based on theoretical 
predictions, particularly our own detailed numerical modeling of the emission in M\,82 and 
NGC\,253 (using a modified version of the GALPROP code: Persic et al. 2008 and Rephaeli et al. 
2010; see also Domingo-Santamar\'\i a \& Torres 2005 for NGC\,253). These analyses do suggest 
that while the relative contribution of the disk is not negligible, it comprises only a small 
fraction due to further steepening of the particle spectra in the disk. However, given the 
uncertainty in the exact values of both the measured and predicted power-law indices, there is 
also an uncertainty in the relative contributions of the SBN and disk.

Our work here improves on our earlier discussion and partial implementation of the radio-based 
method for determining electron and proton energy densities in active regions of SF (Persic \& 
Rephaeli 2010). We do so by accounting for electron radiative losses, by using a more accurate 
calculation of the p/e ratio, and basing our approach on insight gained from our previous 
implementation of a modified version of the GALPROP code that fully accounts for both spectral and 
spatial evolution of proton and electron distributions in the SB region and throughout the disk 
(Persic et al. 2008; Rephaeli et al. 2010). Our general approach and its quantitative implementations 
for predicting the high-energy spectra of the starburst galaxies NGC\,253 and NGC\,3034 have been 
validated by good agreement with $\gamma$-ray measurements (Acero et al. 2009; Acciari et al. 2009; 
Ackermann et al. 2012).

Following on previous work (Beck \& Krause 2005), in a recent paper Lacki \& Beck (2013, hereafter LB13) 
discussed the validity of field-particle equipartition in a SB environment, accounting for secondary 
electrons and strong energy losses. While their main conclusions on deducing $U_{\rm p}$ from radio 
measurements do agree with ours (in both Persic \& Rephaeli 2010 and this work), there are substantial 
differences between our respective treatments. First, we start with the electron spectrum as deduced from 
radio measurements, including the contribution of secondary electrons to the emission. We then use an 
analytically derived primary p/e energy density ratio (using the relevant parameter values) to compute 
$U_{\rm p}$. In contrast, the starting point of the LB13 analysis is the proton energy density, which 
they take to be related to the electron energy density by the same factor, $75$, for all the galaxies in 
their sample. Second, whereas we assume electric neutrality of the accelerated particles to determine the 
primary p/e ratios for different electron and proton spectral indices, LB13 start from an assumed universal 
p/e number density ratio, which they adjust by accounting for the electrons' energy losses. Third, we compute 
the secondary-to-primary electron mumber ratio by using the primary p/e number ratio, the mean proton 
free path in a gas with a given density, and the branching ratios in {\it pp} interactions, whereas LB13 
estimate this ratio by scaling the injection p/e ratio by 1/6 the value of the estimated fraction of 
the proton energy that goes into $\pi^\pm$. These differences in approach and implementation led to 
substantially different results for $U_{\rm p}$ and $B$ (listed in our Table\,2 and their Table\,3). 
Whatever the details, it should be noted that for the nearby starburst galaxies NGC\,253, NGC\,3034, and 
NGC\,4945 our estimated $U_{\rm p}$ and $B$ agree with results of direct measurements. This agreement 
suggests that the equipartition assumption, which we made to derive CR and $B$ energy densities, is 
globally verified in these SBNs, at least as an average property. For NGC\,3034 this suggestion was 
made early on by V\"olk et al. (1990) on the condition that $\zeta(>1\,{\rm GeV}) \sim 10^2$, a 
condition that is consistent with our procedure, see Eq.\,(\ref{eq:zeta3}). 

If the agreement between radio estimates and $\gamma$-ray measurements of $U_{\rm p}$ and B in SBNs is 
further validated and established, the radio method for reliable estimation of proton energy densities 
will be particularly useful for distant ($z \magcir 1$) galaxies whose intense SF is exemplified (albeit 
at lower levels) by the nearby starburst galaxies, but whose faint $\gamma$-ray fluxes are not detectable 
with current or upcoming instruments, whereas their $<$0.1\,mJy radio fluxes are (e.g., Tozzi et al. 2009).

\end{document}